\newcommand{\bea}{\begin{eqnarray}}       
\newcommand{\eea}{\end{eqnarray}}       
\newcommand{\bear}{\begin{eqnarray*}}       
\newcommand{\eear}{\end{eqnarray*}}       

\documentstyle[aps,preprint]{revtex}       
\begin{document}       
       
\draft       
       
\title       
{Interpolation between Hubbard and supersymmetric t-J models. \\
Two-parameter integrable models of correlated electrons.}
       
\author{       
F. C. Alcaraz$^1${\footnote{E-mail address: alcaraz@power.ufscar.br}} 
and R. Z. Bariev$^{1,2}${\footnote {E-mail address: 
rzb@power.ufscar.br}}}       
       
\address{$^1$Departamento de F\'{\i}sica,        
Universidade Federal de S\~ao Carlos, 13565-905, S\~ao Carlos, SP       
Brazil}       
       
\address{$^2$The Kazan Physico-Technical Institute of the Russian        
Academy of Sciences, Kazan 420029, Russia}       
       
\maketitle

\begin{abstract}       
Two new one-dimensional fermionic models depending on two        
independent  parameters are formulated and solved exactly        
by the Bethe-ansatz method.  These models connect continuously the 
integrable Hubbard and supersymmetric t-J models.  
       
\end{abstract}       
       
       
\narrowtext              
The Hubbard model together with the t-J model are the most studied models 
describing strongly correlated electrons. In one dimension they are paradigm 
of exact integrability in the physics of 
 strongly correlated systems. 
In these models 
we have beyond a hopping term t (kinetic energy) a on-site coulomb 
interaction U, in the case of the Hubbard model [1], or a spin-spin 
interaction J, in the case of the t-J model [2-4]. 

An interesting question in the arena of exact integrable models, we want to 
solve in this Letter, concerns the existence of a general exact solvable 
model containing these two well know models as particular cases. 
After the exact solution of these models [1-4], several extensions which 
keep exact integrability were proposed, either by introducing 
correlated hopping terms [5-11], or by including an anisotropy 
(q-deformation) [12-15] (see ref.[16] for a review).
However none of these extensions contains simultaneously the  Hubbard and 
t-J models as particular cases. In this Letter we present two new integrable 
two-parameter models having  this nice property. These models contain as 
particular cases the Hubbard model [1] and the Essler-Korepin-Schoutens 
model [6] as well as its q-deformed versions [12,14,15]. We remind the 
reader that the latter model [6] contains the supersymmetric t-J model
in  a particular sector.

Our starting point is the introduction of a general one-dimensional 
Hamiltonian containing all the possible nearest-neighbour interactions 
appearing in different exactly  
integrable models with four degrees of freedom 
per site. This Hamiltonian thus contains correlated hopping terms  in 
the most general form, spin-spin interactions as in the 
anisotropic version of the t-J model, as well as pair hopping terms and 
three- and four-body static interactions between electrons. The Hamiltonian 
is given by

\bea       
H&=& -\sum_{j=1}^L H_{j,j+1},\nonumber\\       
H_{j,k} &=&\sum_{\alpha(\ne\beta)}(c_{j,\alpha}^+c_{k,\alpha}+h.c.)       
[1 + t_{\alpha 1}n_{j\beta} + t_{\alpha 2}n_{k\beta}+       
t_{\alpha}'n_{j\beta}n_{k\beta}]\nonumber\\       
 &+&\sum_{\alpha(\ne\beta)}(Jc_{j,\alpha}^+c_{k,\beta}^+
 c_{j,\beta}c_{k,\alpha} +      
 V_{\alpha\beta}n_{j,\alpha}n_{k,\beta}) + Un_{j,1}n_{j,2} \nonumber\\       
 &+&t_p(c_{j,1}^+c_{j,2}^+c_{k,2}c_{k,1} + h.c.) +        
 V_3^{(1)}n_{j,2}n_{k,1}n_{k,2} + V_3^{(2)}n_{j,1}n_{k,1}n_{k,2}\nonumber\\       
 &+& V_3^{(3)}n_{j,1}n_{j,2}n_{k,2} + V_3^{(4)}n_{j,1}n_{j,2}n_{k,1}        
 + V_4 n_{j,1}n_{j,2}n_{k,1}n_{k,2};       
\eea       
where $c_{j,\alpha}$ and $n_{j,\alpha} = c_{j,\alpha}^+c_{j\alpha}       
(\alpha =1,2)$ are the standard fermionic and density operators. The        
physical relevance of such Hamiltonian is discussed, e.g. in [17,18].

In (1) we have included a correlated-hopping interaction in its most 
general form, which depends on $t_{\alpha 1},t_{\alpha 2}$ and 
$t_{\alpha}^{'}( \alpha =1,2)$. In the theory of exactly integrable 
systems, models with such kinetic terms were first studied in [5,19] and 
the possible physical relevance of them  is given in [20]. In the 
limit $t_{\alpha \beta} = -t_{\alpha}^{'} = -1$ this term gives a constrained 
hopping term and the condition for integrability gives the anisotropic 
t-J model at 
$J = e^{-\gamma}V_{12} = e^{\gamma}V_{21} = \pm1, t_p = U =        
V_3^{(i)} = V_4 =0$ 
{\footnote { For these parameters the  number of double 
occupied sites are conserved and the model corresponds to that 
introduced in [6]. The t-J 
model is obtained in the sector where there is no double occupied 
sites.} }.  
The Hubbard model is obtained by destroing the 
correlation in the hopping term $(t_{\alpha \beta} = t_{\alpha}^{'} = 0)$ and 
by setting 
 $t_p = J = V_{12} = V_{21} = V_3^{(i)} = V_4 =0$. For the case where $J=0$ 
the conditions for integrability has been investigated in [9,10], and a 
two parameter generalization of the correlated-hopping model has been 
contructed in [9,11]. Recently some one-parameter models with $J \neq 0$  
has been constructed [14,15,21] on the base of solutions of  Yang-Baxter 
equations of vertex models [14,22,23].
In this letter        
we present the results of our investigation on thr integrability conditions 
in the case $J\ne 0, V_{\alpha\alpha} = 0$ and $ t_{\alpha\beta} \ne 1$.

We require the  wavefunctions of the Hamiltonian (1), with n electrons, to be 
given by the Bethe Ansatz 
\bea       
|n> = \sum_{Q} \Psi(r_{Q_1},\alpha_{Q_1};\ldots;r_{Q_n},\alpha_n) 
|r_{Q_1},\ldots,r_{Q_n}> \nonumber \\
\Psi(r_1,\alpha_1;...;r_n,\alpha_n) = \sum_{P} A_{P_1...P_n}^       
{\alpha_{Q_1}...\alpha_{Q_n}}\prod_{j=1}^{n} x_{P_j}^{r_{Q_j}},
\;\;\;\;\;\;\;\;       
x_j = \mbox{exp}(ik_j)       
\eea       
where $Q$ is the permutation of the $n$ particles such that        
$ 1 \le r_{Q_1}\le r_{Q_2}\le ... \le r_{Q_n} \le L$, and $\alpha=1,2$  
denotes  the kind of particles (up or down spin). 
The sum is over all permutations       
 $P =[P_1...P_n]$ of numbers $1,2,...,n$. The coefficients        
$A_{P_1...P_n}^{\alpha_{Q_1}...\alpha_{Q_n}}$ from regions other        
than $R_Q =[r_{Q_1}\le ... \le r_{Q_n}]$ are connected with each other        
by the elements of the two-particle S-matrix       
\bea              
A_{...P_1P_2...}^{...\alpha\beta...} = -              
\sum_{\alpha',\beta' = 1,2}               
S_{\alpha'\beta'}^{\alpha\beta}(k_{P_1},k_{P_2})              
A_{...P_2P_1...}^{...\beta'\alpha'...}      .        
\eea       
As a necessary  condition for integrability of the model       
under consideration, the two-particle scattering matrix has to satisfy       
the Yang-Baxter relations [24,25].       
Although we have not solved this problem in the general case  we were able to 
establish exact integrability of (1) in the two new cases, which we denote 
by models A and B:\\       
A)       
\bea       
t_1 &=&\varepsilon t_2 = t_3 = \varepsilon t_4 =\sin\vartheta; \;\;\;\;       
t_5 = \varepsilon\nonumber\\       
J&=& -\varepsilon t_p = -\frac{\varepsilon}{2}U = V_{12} e^{2\eta} =       
V_{21} e^{-2\eta} = \cos\vartheta;\nonumber\\       
V_{\alpha\alpha} &=&V_{3}^{(1)} = V_{3}^{(2)} = V_{3}^{(3)} = 
V_{3}^{(4)} = V_{4} = 0,       
\eea       
B)       
\bea       
t_1 &=&\varepsilon t_2 = \varepsilon t_3 e^{2\eta} =        
 t_4 e^{-2\eta} =\sin\vartheta; \;\;\;\;       
t_5 = \varepsilon\nonumber\\       
J&=& -\varepsilon t_p = V_{12} e^{2\eta} =       
V_{21} e^{-2\eta} = \cos\vartheta;\nonumber\\       
U &=& 2 t_p + \frac{\sin^2\vartheta}{\cos\vartheta}(e^{\eta} -        
\varepsilon e^{-\eta})^2 \nonumber\\       
 V_{\alpha\alpha} &=&V_{3}^{(2)} = V_{3}^{(4)}= V_{4} =0, \;\; V_{3}^{(1)} 
 = -V_{3}^{(3)} =  V_{12} - V_{21},       
\eea       
where    in (4) and (5) we denote    
\bea      
t_{11} &=& t_4 -1; t_{12} = t_3 - 1; t_{21} = t_1 - 1; t_{22} = t_2 - 1;
\nonumber\\       
t'_1 &=& t_5 - t_3 - t_4 + 1; \;\;\;\; t_2' = t_5 - t_1 - t_2 + 1, \nonumber       
\eea       
$\varepsilon = \pm 1$ and $\vartheta$ and $\eta$ are  free complex 
parameters.
       
For  $\vartheta \rightarrow 0$       
both cases reduce to the anisotropic t-J model  studied in [12],         
which is the generalization of the supersymmetric t-J model [2-4]. More       
exactly, in this limit we obtain the q-deformed  extended Hubbard model [6]. 
Moreover       
from (1,4,5) we see that  the model B with ,   
$\eta = i\vartheta , \varepsilon = + 1$,         and the model A
with   $\eta = 0 ,  \varepsilon = - 1$, 
reduce to       
the  nontrivial q-deformations of the extended Hubbard model considered        
in [14] and [15], respectively. These models have been constructed         
on the base of solution of the 
Yang-Baxter equation for the R-matrix which        
was found by [14,23]. In the        
opposite limit $\vartheta \rightarrow \pi/2$ both models with 
$\varepsilon = 1$ give us the Hubbard model, provide in model A
$\eta = [\ln(U')-\ln(\cos\vartheta)]/2$, and in model B         
$\eta = \frac{1}{2}\sqrt{U|\vartheta - \pi/2|}$.

The non-vanishing elements of the two-particle S-matrix of        
both models satisfy
\bea       
S_{\alpha\alpha}^{\alpha\alpha} &=& 1;\;\;\;       
S_{\alpha\beta}^{\alpha\beta} = S_{\beta\alpha}^{\beta\alpha},\nonumber\\       
S_{\alpha\beta}^{\beta\alpha}(x_1,x_2) S_{\alpha\beta}^{\alpha\beta}(x_2,x_1)
&=&-S_{\beta\alpha}^{\alpha\beta}(x_2,x_1) S_{\alpha\beta}^
{\alpha\beta}(x_1,x_2),       
\eea       
and for the different models are given by \\       
A)       
\bea       
S_{\alpha\beta}^{\alpha\beta}(x_1,x_2) &=& (x_1 -x_2)b_{12}(x_1,x_2)       
/a_1(x_1,x_2);\nonumber\\       
S_{\beta\alpha}^{\alpha\beta}(x_1,x_2) &=& [c_0(x_1,x_2) + b_1(x_1,x_2) x_1 +        
b_2(x_1,x_2) x_2 - g x_1 x_2]/a_1(x_1,x_2);       
\eea       
B)       
\bea       
S_{\alpha\beta}^{\alpha\beta}(x_1,x_2) &=& (x_1 -x_2)b_{12}(x_1,x_2)       
/a_2(x_1,x_2);\nonumber\\       
S_{\beta\alpha}^{\alpha\beta}(x_1,x_2) &=& [c_0(x_1,x_2) + (x_1 e^{-2\eta}       
+ x_2 e^{2\eta})b_{12}(x_1,x_2)]/a_2(x_1,x_2);       
\eea       
where $\alpha < \beta$ and       
\bea       
a_1(x_1,x_2) &=& c_0(x_1,x_2) + [b_1(x_1,x_2) + b_2(x_1,x_2)]x_2 - g x^2_2;       
\nonumber\\       
a_2(x_1,x_2) &=& c_0(x_1,x_2) + (e^{2\eta}+  e^{-2\eta})b_{12}(x_1,x_2) x_2;       
\nonumber\\       
b_1(x_1,x_2) &=& (t_1^2 + \varepsilon J^2 e^{-2\eta})D_{12} + J e^{-2\eta} 
(x_1+x_2);      
\nonumber\\       
b_2(x_1,x_2) &=& (t_1^2 + \varepsilon J^2 e^{2\eta})D_{12} + J e^{2\eta} 
(x_1+x_2);       
\nonumber\\       
b_{12}(x_1,x_2) &=& \varepsilon D_{12} + J(x_1 + x_2);\nonumber\\       
c_0(x_1,x_2) &=& (U - 2 t_p)x_1 x_2 + [t_p D_{12}-x_1 - x_2]D_{12};
\nonumber\\        
D_{12} = 1 &+& x_1 x_2; \;\;\;       
g = \cos\vartheta\sin^2\vartheta(e^{\eta} - \varepsilon e^{-\eta})^2.
\nonumber       
\eea       
       
To complete the proof of the Bethe anzatz (2) we must check the eigenvalue        
equations in the sector where the total number of particles is $n = 3,4$. 
This gives a complicated system of equations. 
A manipulation of this problem on a computer          
gives us the values of the coupling constants $V_3^{(i)}$ and $V_4$        
in eqs. (4) and (5). The periodic boundary conditions on the lattice with 
L sites 
 lead us to the Bethe-ansatz equations. In order        
to obtain these equations we must 
diagonalize the transfer matrix of a related        
inhomogeneous six-vertex 
model with Boltzmann weights (6-8). This latter problem can be        
solved by standard algebraic methods [26]. 
The Bethe-ansatz equations are        
written in terms of       
the variables $x_j$ ($x_j = \mbox{exp}(ik_j)$) and 
additional spin variables       
$x_{\alpha}^{(1)}$.       
       
For both models we have       
\bea       
(x_j)^L = (-1)^{n-1}\prod_{\alpha=1}^m S_{12}^{12}(x_j,x_\alpha^{(1)});       
\nonumber\,\;\;j=1,...,n;\\       
\prod_{j=1}^n S_{12}^{12}(x_j,x_\alpha^{(1)}) =        
\prod_{\beta = 1,\beta\ne\alpha}^m\frac{S_{12}^{12}       
(x_{\beta}^{(1)},x_{\alpha}^{(1)})}{S_{12}^{12}       
(x_{\alpha}^{(1)},x_{\beta}^{(1)})},\;\;\;j=1,...,m;       
\eea       
where $m \leq L$ is the number of particles with up spins. The eigenenergies of        
the system are given by       
\bea       
E = -\sum_{j=1}^n(x_j + x_j^{-1}).       
\eea       

An important step toward the solution of integrable models, in the 
thermodynamic limit, is the definition of new variables $\lambda(x_j)=
\lambda_j$, in terms of which $S_{12}^{12}(x_i,x_j)$ becomes a function only 
of the difference $\lambda_i - \lambda_j$. The corresponding integral 
equation derived from (9) will then have difference kernels. 
Following Baxter [25], we introduce a function       
\bea       
\lambda(x_1,x_2) = \frac{1}{2}\ln\frac{1 + e^{-2r} \Phi(x_1,x_2)}       
{1 + e^{2r} \Phi(x_1,x_2)}, \;\;\; \Phi(x_1,x_2) = S_{12}^{12}(x_1,x_2);       
\eea       
where r is the Baxter parameter, conveniently chosen for our models 
\bea            
\cosh2r = \cases{              
\varepsilon t_1^2 + J^2\cosh2\eta,& \mbox{for model A}\cr              
\cosh2\eta,& \mbox{for model B}\cr}   .           
\eea            
       
It follows (see  [25])  the function $\lambda(x_1,x_2)$        
have the nice  property       
\bea       
\lambda(x_1,x_3) = \lambda(x_1,x_2) + \lambda(x_2,x_3),       
\eea       
which imply 
 \bea       
\lambda(x_1,x_2) = \lambda(x_1) - \lambda(x_2).       
\eea       
Using (11) and (14) we rewrite the  Bethe-ansatz equations in the        
difference form       
difference form       
\bea       
(x_j)^L = \prod_{\alpha=1}^m \frac{\sinh(\lambda_j -       
\lambda_{\alpha}^{(1)} -r)}{\sinh(\lambda_j -       
\lambda_{\alpha}^{(1)} +r)}, \;\;\;\;\;\;j=1,...,n;       
\nonumber\\       
 \prod_{j =1}^n \frac{\sinh(\lambda_j -       
\lambda_{\alpha}^{(1)} -r)}{\sinh(\lambda_j -       
\lambda_{\alpha}^{(1)} +r)} = -\prod_{\beta =1}^m        
\frac{\sinh(\lambda_{\beta}^{(1)} -       
\lambda_{\alpha}^{(1)} - 2r)}{\sinh(\lambda_{\beta}^{(1)} -       
\lambda_{\alpha}^{(1)} + 2r)},\;\;\;\alpha=1,...,m       .
\eea       
where $\alpha$ and $\beta$ are the arbitrary values.  For example,       
we may choose $\alpha = 0$ and  $\beta = r$ for our convenience. The       
function $\Phi(x,0)$ has the same form for both models, namely  
\bea       
\lambda(x) = \frac{1}{2}\ln\frac{1 + e^{-2r} \Phi(x,0)}       
{1 + e^{2r} \Phi(x,0)} + r, \;\;\;\;\;\;\;\;\Phi(x,0) =  
\frac{-x(\varepsilon + Jx)}{(\varepsilon J + x)} .       
\eea        
The inversion of (16) gives us        
\bea       
x = \frac{-J^{-1}\cosh\lambda \sinh r \pm \sqrt{\sinh^2\lambda\cosh^2 r +        
\cosh^2\lambda\sinh^2 r\tan^2\vartheta}}{\sinh(\lambda + r)};  \;\;       
\varepsilon = +1; \nonumber\\       
x = \frac{J^{-1}\sinh\lambda \cosh r \pm \sqrt{\cosh^2\lambda\sinh^2 r +        
\sinh^2\lambda\cosh^2 r\tan^2\vartheta}}{\sinh(\lambda + r)}; \;\;        
\varepsilon = -1.        
\eea       
It is clear from (16) that the Bethe-ansatz equations have the .
same form for        .
both models at the same values of the parameters $r$ and $\vartheta$.       .

Let us consider the Bethe-ansatz equations in some limiting cases. At 
$\cos\vartheta \rightarrow 1$ we obtain $x_j = \sinh(\lambda_j -r)/
\sinh(\lambda_j +r)$ and (15) gives us the Bethe-ansatz equations of the 
anisotropic supersymmetric t-J model, with anisotropy r [12]. In our 
derivation of (15) we assume the amplitudes in the eigenfunctions,   
corresponding to double site occupations, are related with those with single 
occupancy. Strictly at  $\cos \vartheta =1$, this assumption is not valid, 
unless there is no double occupancy as in the t-J model, and we should  
 restrict  $m \leq L/2$ in (15). 

In the limiting case of model B with $\varepsilon =1$, where 
$\cos \vartheta  \rightarrow 0, \eta \rightarrow 0$, with 
$ U = 4 \eta^2/\cos  \vartheta $ fixed we obtain from (5) the Hubbard 
model with on-site interaction $\tilde U = U$. The relation (12) gives us 
$r = \sqrt{U\cos (\vartheta) }/2$ and by choosing 
$\lambda_j = i(\pi/2 - 2\sin k\sqrt{\cos (\vartheta) /U})$, 
$\lambda_j^{(1)} = i(\pi/2 - 2\Lambda_j\sqrt{\cos(\vartheta)/U})$ we obtain 
the Bethe-ansatz equations of the Hubbard model [1]
\bea
e^{ik_jL} = \prod _{\alpha=1}^{m} \frac{\sin k_j - \Lambda_j - i\tilde U/4} 
{\sin k_j - \Lambda_j + i\tilde U/4}, \;\;\;j=1,...,n\\
\prod_{j=1}^n \frac{\sin k_j - \Lambda_\alpha + i\tilde U/4}
{\sin k_j - \Lambda_\alpha - i\tilde U/4} = 
-\prod_{\beta=1}^n \frac {\Lambda_\beta -\Lambda_\alpha + i \tilde U/2}
{\Lambda_\beta - \Lambda_\alpha - i \tilde U/2}\;\;\;\;j=1,...,m.
\eea
The Hubbard limit can also be obtained  in the limiting case of model A 
with $\varepsilon =1$ where $\cos \vartheta  \rightarrow 0, \eta 
\rightarrow 0$, but 
$\tilde U = V_{21} = e^{2\eta}\cos(\vartheta)/2$  kept fixed.
In this case we see from (5) that shifting 
$c_{j,2} \rightarrow c_{j-1,2}$, we recover the Hubbard model with on-site 
interactions $\tilde U = V_{21}$. The Bethe-ansatz equations (18) are obtained 
from (15) by choosing 
$\lambda_j = i(\pi/2 - 2e^{-\eta}\sin k_j)$ and 
$\lambda_{\alpha}^{(1)} = i(\pi/2 -2e^{-\eta}\Lambda_\alpha)$. 

It is also interesting to observe that  rational Bethe-ansatz equations 
can also be obtained for both models in the limit where $r \rightarrow 0$ or 
$r \rightarrow i\pi/2$. We should remark that even in this case we obtain 
new integrable quantum chains. For example at  $r \rightarrow 0$ we can 
rewrite (15) as 
\bea
e^{ik_jL} = \prod_{\alpha =1}^m \frac{\lambda_j -\Lambda_\alpha +\frac{i}{2}}
{\lambda_j -\Lambda_\alpha - \frac{i}{2}}, \;\;\;j =1,...,n,\\
\prod_{j=1}^m \frac{\lambda_j - \Lambda_\alpha - \frac{i}{2}} 
{\lambda_j - \Lambda_\alpha + \frac{i}{2}} = 
-\prod_{\beta=1}^n \frac{\Lambda_\beta - \Lambda_\alpha - i}
{\Lambda_\beta - \Lambda_\alpha + i}, \;\;\; \alpha = 1,...,m;
\eea
where 
\bea
\lambda_j = \frac{1}{4}[(J^{-1} + 1) \cot(k_j/2) + 
(J^{-1} -1) \tan(k_j/2) ],\;\;\; \mbox{for} \;\;\; \varepsilon =+1,
\eea
and  
\bea
\lambda_j = [(J^{-1} + 1) \tan(k_j/2) + 
(J^{-1} -1) \cot(k_j/2) ]^{-1} 
,\;\;\; \mbox{for} \;\;\; \varepsilon =-1.
\eea
These solutions correspond to the model A at $\eta =0, \varepsilon = +1$
and at $\cos\vartheta \cosh\eta = \pm 1, \varepsilon = -1$, and to
model B at $\eta = 0$ for both signs of $\varepsilon$. 

To summarize we have presented two new two-parameter integrable models that 
generalize the Hubbard and supersymmetric t-J models, and  derived their 
Bethe-ansatz equations through the coordinate Bethe-ansatz method.
Our results certainly motivate subsequent studies. One of them is the 
calculation of the phase diagram and critical exponents for arbitrary 
values of $\eta$ and $\vartheta$. Another interesting point raised by the 
present work, is the possible existence of a  
 generalized R-matrix that 
reproduces that of the Hubbard model [27] at special points. 
It will be also worthwhile to generalize the model (1) for the
case $\alpha > 2$ and to construct, in such way, the quite interesting
Hamiltonian of multy-color Hubbard model [16,28].
       
We thank  M. J. Martins and V. Rittenberg for useful  discussions and  
also   A. Lima-Santos and J. de Luca for interesting conversations. 
This work was supported in part by Conselho Nacional               
de Desenvolvimento        
Cient\'{\i}fico e Tecnol\'ogico  - CNPq - Brazil.

\end{document}